\documentclass[aps,prl,superscriptaddress,twocolumn]{revtex4}
\usepackage{graphicx}
\usepackage{setspace}
\usepackage{rotating}
\begin{document}

\def\Eint{{E_{\rm int}}}
\def\Eads{{E_{\rm ads}}}
\def\PBEkappa{{PBE$\kappa$=1}}
\def\PBEalpha{{PBE$\alpha$=2}}
\def\DeltaHB{{$\Delta^{\rm H-D}$}}
\title{Chemical accuracy for the van der Waals density functional}
\author{Ji\v{r}\'\i\ Klime\v{s}}
\affiliation{London Centre for Nanotechnology and Department of Chemistry, University College London, London, WC1E 6BT, UK}
\author{David R. Bowler}
\affiliation{London Centre for Nanotechnology and Department of Physics and Astronomy, University College London, London, WC1E 6BT, UK}
\author{Angelos Michaelides}
\email{angelos.michaelides@ucl.ac.uk}
\affiliation{London Centre for Nanotechnology and Department of Chemistry, University College London, London, WC1E 6BT, UK}


\date{\today}

\begin{abstract}

The non-local van der Waals density functional (vdW-DF) of Dion~{\it et al.} [Phys. Rev. Lett. {\bf 92}, 246401 (2004)]
is a very promising scheme for the efficient treatment of dispersion bonded systems.
We show here that the accuracy of vdW-DF can be dramatically improved both for dispersion and hydrogen bonded complexes
through the judicious selection of its underlying exchange functional. 
New and published exchange functionals are identified that deliver much better than chemical accuracy from vdW-DF
for the S22 benchmark set of weakly interacting dimers and for water clusters.
Improved performance for the adsorption of water on salt is also obtained.

\end{abstract}

\maketitle

London dispersion interactions are ubiquitous in nature contributing to the binding of biomolecules such as DNA, molecular
crystals, and molecules on surfaces. 
The accurate description of dispersion, which often occurs in conjunction
with hydrogen bonds, is a major challenge for many electronic structure theories. 
Density functional theory (DFT), the most widely used electronic structure theory, often doesn't meet this challenge.
Indeed, it is well-established that popular generalized gradient approximation (GGA) or hybrid exchange-correlation functionals
are inadequate for the description of dispersion interactions.
Many schemes have been developed that allow dispersion to be accounted for within DFT in a more or less approximate 
manner (see, e.g. ~\cite{dion2004,lilienfeld2004,antony2006,becke2007,grimme2007,sato2007,tkatchenko2009}). 
One of the most promising and rigorous methods is the non-local 
van der Waals density functional (vdW-DF) of Langreth and Lundqvist and co--workers~\cite{dion2004}.

In vdW-DF the non-local correlation is calculated so that the exchange-correlation energy takes the form 
\begin{equation}
E_{\rm xc}=E_{\rm x}^{\rm GGA}+E_{\rm c}^{\rm LDA}+E_{\rm c}^{\rm nl}\,,
\label{eq_func}
\end{equation}
where $E_{\rm x}^{\rm GGA}$ is the GGA exchange energy.
In the original vdW-DF this is obtained with the revised version of the 
Perdew, Burke, and Ernzerhof (PBE)~\cite{perdew1996} functional 
from Zhang and Yang (revPBE)~\cite{zhang1998}. 
$E_{\rm c}^{\rm LDA}$ accounts
for the local correlation energy obtained within the local density approximation (LDA), 
and $E_{\rm c}^{\rm nl}$ is the non-local correlation energy.
The formula for $E_{\rm c}^{\rm nl}$ is based on electron densities interacting via a model response function, 
the particular form of which is still a subject of research~\cite{vydrov2009}.
The vdW-DF has been applied to a wide variety of systems where dispersion is important (see ref.~\cite{langreth2009} for a review) 
and recent algorithmic developments~\cite{soler2009} have made it only marginally more 
computationally expensive than a regular GGA. 
However, in many important circumstances the current vdW-DF is simply not accurate enough. 
For example, for the S22 dataset~\cite{jurecka2006} 
(a set of 22 weakly interacting dimers mostly of biological importance)
it yields a mean absolute deviation (MAD) of $\sim$60~meV~\cite{gulans2009}
compared to coupled cluster reference data. 
This is outside the so-called ``chemical accuracy" of 1~kcal/mol or $\sim$43~meV and 
inferior performance to other DFT-based dispersion correction schemes~\cite{antony2006, grimme2007,tkatchenko2009}.
% such as the methods of Tkatchenko and Scheffler 
%(MAD=13~meV for S22)~\cite{tkatchenko2009} or Grimme (MAD=15~meV)~\cite{antony2006, grimme2007}. 
Water clusters, important for atmospheric chemistry and liquid water,
are another example where vdW-DF substantially underbinds (by $\sim$20\% compared to accurate reference data)
and in terms of absolute dissociation energies is worse than a regular GGA such as PBE~\cite{kelkkanen2009,santra2008}.

Recognizing that the interaction energies obtained 
with vdW-DF depend on the exchange functional incorporated within it~\cite{dion2004,thonhauser2006,gulans2009},
we aimed to improve vdW-DF 
by exploring and developing alternative exchange functionals to the original revPBE.
We take a pragmatic approach, we use $E_{\rm c}^{\rm nl}$ in its regular form and search 
for an exchange functional that combines with it to give precise energies for a wide range of systems.
To this end we first use the S22 dataset since it includes a variety of weakly bonded dimers for which accurate interaction
energies and structures have been established~\cite{jurecka2006} and so provides a tough test for molecular simulation methods.
Following this we test our methods on two complex systems where dispersion interactions are crucial: water hexamers
and water adsorbed on NaCl(001).
From these studies we propose three new exchange functionals, 
that when incorporated within vdW-DF offer vastly improved interaction energies compared to those from the original vdW-DF. 
The new functionals, which are easy to implement and come at no extra cost, make
vdW-DF competitive with all other DFT-based methods for the treatment of weak interactions. 
We hope that this study lays the foundations for further improvements of vdW-DF and enables more accurate treatments of dispersion and hydrogen bonded systems, for example, liquid water and ice.

%Technical details
Throughout, we calculate the vdW-DF energies non-self-consistently in two steps. 
First, VASP 5.2 \cite{kresse1993,kresse1996}
calculations with a given exchange functional \cite{new_xc_note} and PBE correlation functional are performed. 
Second, the VASP electron density is used to determine the vdW correction 
%$\Delta E^{\rm vdW}$ to the energy according to 
%$\Delta E^{\rm vdW}=-E_{\rm c}^{\rm PBE}+E_{\rm c}^{\rm LDA}+E_{\rm c}^{\rm nl}$ 
using JuNoLo~\cite{lazic2008b}.
We find that the magnitude of the vdW correction is rather insensitive to the underlying density used~\cite{foot_dens}.
Therefore, density from B86 (exchange) and PBE (correlation) calculations was used for all functionals except PBE and revPBE,
where density from the respective exchange-correlation functional was used.
Care was taken with the VASP calculations to ensure that converged energies were obtained, which involved the use
of hard projector-augmented wave (PAW)~\cite{blochl1994, kresse1999} potentials, an 800 to 1000~eV cut-off, dipole corrections, 
and 20 to 25~\AA$^3$ unit cells. 
Since the efficient self-consistent calculation of vdW-DF energies has only very recently become possible~\cite{soler2009}, we checked 
at the latter stages of this study how the
non-self-consistent and self-consistent interaction energies differ with the grid-based GPAW code~\cite{mortensen2005}. 
For the S22 dataset the non-self-consistent and self-consistent interaction energies are
within 1.5~meV, except for the large dispersion bonded dimers (dimers 11--15) where the differences are $\le$4~meV. 
For the water hexamers the non-self-consistent and self-consistent interaction energies are within $\sim$2~meV.

%Results

\begin{figure}[h]
%\centerline{\includegraphics[width=13cm,]{graph/xc_diff/new_xc}}
\centerline{\includegraphics[width=5.5cm,angle=-90]{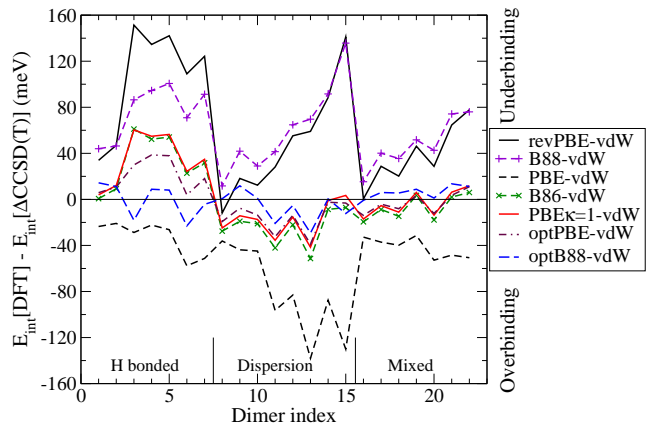}}
\caption{Differences in interaction energies for vdW-DF ($E_{\rm int}$[DFT]) with various exchange functionals
from the CCSD(T) reference data ($E_{\rm int}$[$\Delta$CCSD(T)])~\cite{jurecka2006}. We show data for revPBE, B88, PBE, B86 and three new exchange functionals ``\PBEkappa", ``optPBE", and ``optB88".
%The S22 set comprises the following dimers: (1) ammonia; (2) water; (3) formic acid; (4) formamide;
%(5) uracil; (6) 2-pyridoxine $\cdot$ 2-aminopyridine; (7) adenine $\cdot$ thymine, held by H bonds;
%(8) methane; (9) ethene; (10) benzene $\cdot$ methane; (11) slip-parallel benzene dimer; (12) pyrazine; 
%(13) stacked uracil dimer; (14) stacked indole benzene dimer; (15) stacked adenine thymine dimer, held predominantly by dispersion;
%(16) ethene $\cdot$ ethine; (17) benzene $\cdot$ water; (18) benzene $\cdot$ ammonia; (19) benzene $\cdot$ hydrogen cyanide;
%(20) T-shaped benzene dimer; (21) T-shaped indole benzene dimer; and (22) phenol.
}
\label{fig_diff}
\end{figure}

Let us first examine the results for the standard form of the vdW-DF where revPBE exchange is used.
Throughout, we denote a combination of an exchange functional X with vdW correlation as X-vdW,
hence we refer to vdW-DF as revPBE-vdW.
The differences in the revPBE-vdW and reference interaction energies 
for each of the dimers in the S22 dataset are shown in Fig.~\ref{fig_diff}. 
%The reference interaction energies for the S22 dataset
%are taken from~\cite{jurecka2006} where they were computed at the $\Delta$CCSD(T)/CBS level.
One can see from Fig.~\ref{fig_diff} that with revPBE-vdW
most of the dimers are substantially underbound. 
The MAD is 65~meV (Table~\ref{tab_mad}), which is in good agreement with the MAD of 60~meV
in Gulans~{\it et al.}~\cite{gulans2009}.
% and qualitatively agree with the results published 
%for some of the dimers by Cooper~{\it et al.}~\cite{cooper2008}.
%
The errors for the individual hydrogen bonded (HB), dispersion bonded (DB), and mixed dispersion and hydrogen bonded (MB) subsets
are all quite large at 106, 52, and 38~meV, respectively. 
Further, this functional yields a very large ``Range" of errors (i.e., the difference between the largest and smallest
errors) of 163~meV and does not provide a good balance between H bonding and dispersion. This
is shown by \DeltaHB\ which 
%the ``Range" column gives the difference of the largest and the smallest error, which for this functional is 163~meV.
gives the difference between the mean deviations (MD)
of the HB and DB subsets (i.e., \DeltaHB$=$MD(HB)$-$MD(DB)). The smaller this quantity,  
the better the balance between the different types of bonding. 
The rather large value of \DeltaHB\ for revPBE-vdW of 57~meV reveals that  
on average the HB dimers are underbound compared to the DB dimers.
% which can be seen also in Fig.~\ref{fig_diff}. 
%This difference should be as small as possible 
%to describe accurately the balance between different types of bonding.
Therefore, the original vdW-DF does not deliver
chemical accuracy for either systems held by dispersion or H bonds,
it yields a large range of errors, and on average underbinds the H bonded 
compared to the dispersion bonded systems.

\begin{table}[h]
\caption{Mean absolute deviations from the reference data~\cite{jurecka2006} for the
S22 set for vdW-DF with various exchange functionals (``Method"). MADs are given for the whole set
(``MAD Total''), the hydrogen bonded (``MAD HB"), dispersion bonded (``MAD DB"), and mixed bonding
subsets (``MAD MB"). For each functional we also report the difference between the largest and smallest deviations from the reference
data (``Range") and the difference in the mean deviations of hydrogen bonded and dispersion bonded subsets (\DeltaHB). The new functionals introduced in this study are in the last three rows. All values are in meV.  }
\label{tab_mad}
\centerline{
\begin{tabular}{lcccccc}
\hline\hline
Method  &     \multicolumn{4}{c}{MAD} & Range & \DeltaHB \\     
       & Total & HB & DB &  MB &    &  \\
%\hline
%PBE (no-HB)& 117& 45& 207& 87&466 & -- \\
\hline
revPBE-vdW  &65 &106&52 &38 & 163& 57\\
\hline
%published
%PBEsol-vdW & 115& 124& 140& 77& 233& 16\\
B88-vdW  & 62 & 76 & 61 & 48 & 124 & 16\\
%RPBE-vdW & 54& 102& 35& 26& 185& 75\\
PBE-vdW      & 54& 33& 83& 42& 117& 50\\
%PW86-vdW & 41& 14& 73& 31& 130& 61\\
%B86MGC-vdW  &28 &10 &49 & 20& 97&44 \\
%\PBEalpha-vdW & 25& 26& 34& 13&116 &59 \\
B86-vdW     & 23& 33& 25& 10& 112& 58\\
\hline
%devised
%PBEfix0.5-vdW & 24& 35& 26& 11& 117& 61\\
\PBEkappa-vdW & 21& 35&19 &10 &102 & 54\\
%rRPBE-vdW &20 &29 &22 &9 &99 &51 \\
optPBE-vdW & 15& 21& 16& 8& 77& 37\\
optB88-vdW & 10 &13& 10 &7 &44 & 6 \\
\hline\hline
\end{tabular}}
\end{table}

%previously published functionals
%To understand the poor performance of the revPBE based vdW-DF and in the hope of improving matters, several other
%GGA exchange functionals were tested with vdW-DF, i.e., alternatives to revPBE were considered for $E_{\rm x}^{\rm GGA}$ in (\ref{eq_func}). 
We now consider alternatives to revPBE for $E_{\rm x}^{\rm GGA}$ in (\ref{eq_func}), discussing the results from just a few of the most interesting functionals,
namely PBE, Becke86 (B86)~\cite{becke1986}, and 
Becke88 (B88)~\cite{becke1988}.
As can be seen from Table~\ref{tab_mad} all functionals yield smaller MADs than revPBE-vdW
and overall the performance is qualitatively different. For example, in contrast to revPBE-vdW, PBE-vdW systematically
overbinds the dimers. Although the MAD of B88-vdW is only marginally smaller
(62~meV) than that of revPBE-vdW, the \DeltaHB\ is reduced from 57 to 16~meV. 
And of most interest, B86-vdW, yields an overall MAD of just 23~meV. 
This is the lowest MAD obtained from all published functionals considered and
a substantial improvement over revPBE.
%(number 8) is off by $-28$~meV (120\% relative error); overbinding of this dimer seems to be a common drawback for the 
%exchange functionals with the PBE form. 
%Of the other functionals considered B86MGC-vdW offers rather good performance with a MAD of 28~meV.
%PBE-vdW systematically overbinds the dimers in contrast to revPBE which underbinds all of them (except for the methane dimer).
%This contrasting performance of PBE-vdW and revPBE-vdW, which was also observed recently by Gulans~{\it et al.}, provides
%the necessary physical insight to identify improved exchange functionals.
%Although B88 underbinds the DB dimers like revPBE does, it binds strongly the HB pairs so that \DeltaHB\  
%is only 17~meV compared to \DeltaHB=57~meV for revPBE. 
%This corresponds to the fact that although the B88 enhancement factor has similar values to B86 
%for $s< 2 $ it keeps rising after that point which causes destabilization of the DB complexes compared to separated monomers.
%Interestingly, although the B86MGC exchange functional was previously identified to give similar dissociation curves 
%to HF for rare-gas dimers~\cite{lacks1993}, B86MGC-vdW overbinds the DB dimers.
%However, this may be in concert with the fact that the current form of the vdW functional does not perform 
%well when paired with the HF exchange~\cite{puzder2006,vydrov2009}. 
%going from PBE to R and revPBE

\begin{figure}[h]
%\centerline{\includegraphics[width=13cm]{enh_new}}
\centerline{\includegraphics[width=5.5cm,angle=-90]{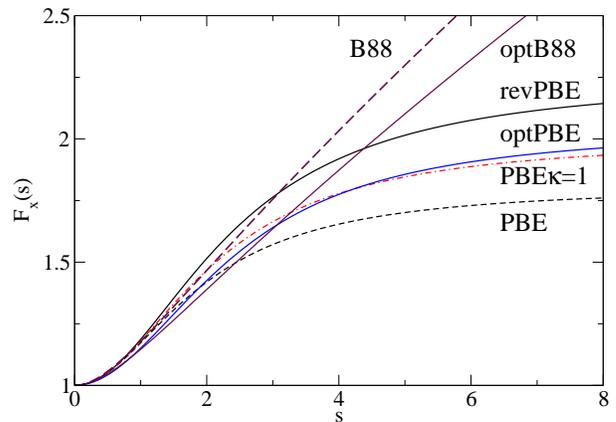}}
\caption{Enhancement factors of some of the exchange functionals discussed. PBE, revPBE, and the new \PBEkappa\ 
all have the form given by~(\ref{eq_pbe}) but differ in the value of $\kappa$.
optPBE is a combination of PBE and RPBE with parameters optimized for the S22 dataset.
Similarly optB88 has the form given by~(\ref{eq_b88}) and is again parameterized for the S22 dataset.
The enhancement factor of B86 is very similar to \PBEkappa\ and is not shown for clarity.}
\label{fig_enh}
\end{figure}

Can the errors on the S22 set be further reduced?
The contrasting performance of PBE-vdW and revPBE-vdW, which was also observed by Gulans~{\it et al.}~\cite{gulans2009}, provides
the necessary physical insight to identify improved exchange functionals.
As we know, the GGA exchange energy density is given
by $\varepsilon_{\rm x}(n,s)=\varepsilon^{\rm LDA}_{\rm x}(n) F_{\rm x}(s)$, where
$\varepsilon^{\rm LDA}_{\rm x}(n)$ is the LDA exchange energy density and $F_{\rm x}(s)$
is the enhancement factor that depends on the reduced density gradient $s=|\nabla n|/2(3\pi^2)^{1/3}n^{4/3}$.
The enhancement factors of PBE and revPBE have the same form:
\begin{equation}
F^{\rm PBE}_{\rm x}(s) =1+\kappa-\kappa/(1+\mu s^2/\kappa)\,.
\label{eq_pbe}
\end{equation}
%with %the parameter $\mu=0.21951$ is the same, they differ in the value of the parameter $\kappa$.
The parameter $\mu$ is also the same and so the functionals differ only in the value of the parameter $\kappa$.
%The revPBE has a larger value ($\kappa^{\rm revPBE}$=1.245) 
%than PBE ($\kappa^{\rm PBE}$=0.804) which causes it to prefer dimerization less than PBE does. 
%
revPBE has a larger value of $\kappa$ than PBE ($\kappa^{\rm revPBE}$=1.245, $\kappa^{\rm PBE}$=0.804),
which causes $F_{\rm x}$ to rise more rapidly with revPBE than PBE (see Fig.~\ref{fig_enh} to 
see the enhancement factors).
As a consequence, regions with large reduced density gradients are stabilised more with revPBE than PBE, which 
%
%This favors monomers over dimers and as a consequence 
in turn leads to weaker 
interactions with revPBE (see ref.~\cite{hammer1999} for a more detailed discussion on this issue).
%a result there is a smaller exchange contribution to binding between the dimers from revPBE compared to PBE.
%
Therefore, in principle, a simple strategy for obtaining improved interaction energies is to identify an exchange functional intermediate between PBE and revPBE.  
To this end we varied $\kappa$ from the PBE to revPBE values (in 0.05 increments) and calculated interaction energies within vdW-DF for 
the complete S22 dataset. A value of $\kappa$=1.00 resulted in the smallest MAD of only 21~meV. 
We dub this new exchange functional ``\PBEkappa"~\cite{foot_alpha}. 
%For any DFT code in which PBE is implemented it is obviously trivial to implement
%\PBEkappa~\cite{foot_alpha}. 
%sought functionals that
%ombine the PBE and RPBE~\cite{hammer1999} enhancement factors and optimized the parameters $\mu$, $\kappa$, and the mixing ratio.
Pushing the PBE-style (i.e., PBE and its various revised forms) functionals yet \hbox{further} we varied $\mu$ and $\kappa$, and also
considered other forms of the enhancement factor. After optimization we obtained
an exchange functional ``optPBE" that yielded a MAD of only 15~meV. 
This functional turned out to be a 95\% PBE and 5\% RPBE~\cite{hammer1999} combination with
$\mu=0.176$, and $\kappa=1.05$.
%, and 0.95 of PBE-like exchange and 0.05 of RPBE like exchange is combined. 
The enhancement factor is shown in Fig.~\ref{fig_enh}.

The two PBE-style functionals introduced above offer substantial improvements over revPBE. However, they still
exhibit large errors in \DeltaHB\ and overbind the 
%Although all of the statistical parameters are improved by using this functional compared to, e.g., \PBEkappa, 
%one can see from Figure~\ref{fig_diff} that 
%the systematic errors: large \DeltaHB\ and overbinding of the 
methane dimer (dimer 8, by 25~meV or 108\% with \PBEkappa\ and 19~meV or 85\% with optPBE).
Since B88 is free from these deficiencies we explored optimised versions of it.
The B88 exchange enhancement factor can be written as
\begin{equation}
F^{\rm B88}_{\rm x}(s)=1+\mu s^2/(1+\beta s\, {\rm arcsinh}(c s))\,,
\label{eq_b88}
\end{equation}
where $c=2^{4/3}(3\pi^2)^{1/3}$, $\mu\cong0.2743$, and $\beta= 9\mu(6/\pi)^{1/3}/(2c)$. 
As B88 underbinds the dimers, we modified the ratio $\mu/\beta$ to lead to increased binding,
resulting in an optimal  $\mu/\beta$ of $1.2$ and a $\mu$ of $0.22$.
The new exchange functional, which we dub ``optB88'', yields a
%
%There are effectively two parameters present in this functional: the first
%one $\mu$ controls the behavior for small $s$, the other $\beta$ is related to the slope of the enhancement factor. We set $\mu=0.22$,  
%so that for small $s$: $F_{\rm s}(s)=1+0.22s^2$ (which is a similar behavior to PBE, $\mu$=0.21951), 
%so that for small $s$ the enhancement factor is similar to PBE,
%and adjusted $\beta$ so that a small MAD is obtained. 
%Aiming for a smaller
%Since the B88 gives repulsive behavior, we decreased the slope by changing 
%the $\mu/\beta$ ratio from the B88 value of $\sim$1.4 to 1.2. 
%The final form is then 
%\begin{equation}
%F_{\rm x}(s)=1+\mu s^2/(1+\beta s\, {\rm arcsinh}(c s))\,,
%\label{eq_b88}
%\end{equation}
%with $\mu$=0.22, $\mu/\beta$=1.2, and $c=2^{4/3}(3\pi)^{1/3}$.
%As can be seen from Fig.~\ref{fig_enh}, the enhancement factor is lower than the PBE one for $s$ below 2 but rises over revPBE for s$>$4.
%we obtain a functional that we dub ``optB88". This functional 
%
MAD of only 10~meV, an accurate binding energy for the methane dimer, and similar 
mean deviations for all three subsets.  
Of all the functionals considered, optB88 is the most
accurate for the S22 dataset.

%compare to others
%Let us now compare our results to the data previously published in the literature for different van der Waals correction schemes.
%If we consider the published MADs for the LC-BOP+ALL (20~meV), a method based on a range separated hybrid and a
%simpler version of vdW-DF~\cite{sato2007}, 
%B97-D (15~meV)~\cite{antony2006, grimme2007}, and a recently proposed
%algorithm from Tkatchenko and Scheffler (13~meV)~\cite{tkatchenko2009}, we see that our vdW functional-compatible 
%exchange functional gives better results than these approaches.
%Moreover, all these schemes need to apply a damping function on the vdW correction which makes them less transferable 
%and some of them have been obtained with non-converged basis set.

We now consider whether the improved performance
of vdW-DF carries over to other systems and start with gas phase water hexamers. 
Hexamers are interesting for many reasons (e.g. they are key constituents of the condensed phases of water) and not least because they provide a tough test for DFT~\cite{santra2008}. 
In particular many functionals (including those
widely used to study water such as PBE and BLYP) incorrectly predict that a ``cyclic'' or ``book" cluster has the lowest energy,
in contrast to coupled cluster which favors a ``prism'' structure.
The energy differences between the various isomers are very small ($\leq10$ meV) and only when dispersion is taken into account is the correct energy ordering recovered~\cite{santra2008}. 
%
%Whereas the vdW-DF (with revPBE exchange) improves the relative energy ordering of the low energy isomers~\cite{kelkkanen2009}, the 
%absolute dissociation energies are underestimated by $\ge$50~meV/H$_{\rm 2}$O. 
%
We have tested the various functionals discussed above on the water hexamers  and the results are reported in Table~\ref{tab_hex}. 
Compared to pure PBE the vdW-DF improves the relative energies of the hexamers and B86-vdW and the three new functionals yield considerably
improved absolute dissociation energies over revPBE-vdW~\cite{kelkkanen2009}. 
In particular the dissociation energies for optPBE-vdW are essentially identical 
to those obtained with $\Delta$CCSD(T). 
The fact that there is now a functional which predicts both accurate 
absolute and relative energies for water hexamers is very encouraging and makes optPBE-vdW an interesting prospect for
condensed phase simulations of water.
Finally we note that for one of the new exchange functionals (optB88-vdW) we have performed self-consistent geometry optimizations
which resulted in only slightly different dissociation energies (net differences of $\sim$7 meV, Table II) and, 
moreover, very similar geometries \cite{geom_note}.

\begin{table}[h]
\caption{Dissociation energies (meV/H$_2$O) for four low energy isomers of the water hexamer 
calculated using $\Delta$CCSD(T), PBE (with no vdW correction), and the vdW-DF with various exchange functionals. 
Unless indicated otherwise MP2 geometries from ref.~\cite{santra2008} were used. The $\Delta$CCSD(T) data in this table was computed as part
of this study in the standard way, i.e., the MP2$-$CCSD(T) difference at the triple zeta level
(aug-cc-pvtz basis set) was added to the MP2 complete basis set dissociation energies.}
\label{tab_hex}
\centerline{
\begin{tabular}{lcccc}
\hline\hline
%Method  &     \multicolumn{4}{c}{MAD} & Range & MD(HB) $-$ MD(DB)\\     
       & Prism & Cage & Book &  Cyclic  \\
%\hline
\hline
%$\Delta$CCSD(T) & -332& -330&-327 &-319 \\
$\Delta$CCSD(T) & $-334$&$ -332$&$ -329$&$ -321$\\
%old revPBE-vdW  &-276 &-276&-277 &-273 \\
PBE & $-334$ & $-336$ & $-343$ & $-341$ \\
revPBE-vdW\footnote{From ref. \cite{kelkkanen2009}}  &$-280$ &$-279$&$-277$ &$-269$ \\
%revPBE-vdW(ref X)& -280& -279&-277 & -269\\  
%\hline
%published
%PBEsol-vdW & 115& 124& 140& 77& 233& 16\\
%bad B88-vdW  & -293 & -295 & -295 & -290 \\
B88-vdW  & $-286$ &$ -287$ &$ -288$ &$ -282$ \\
%RPBE-vdW &-281 &-281 & -282& -277\\
%old PBE-dW      &-387 &-385 & -378& -365\\
PBE-vdW      &$-380$ &$-378$ &$ -372$&$ -358$\\
%PW86-vdW & 41& 14& 73& 31& 130& 61\\
%B86MGC-vdW  & -364& -362& -357& -345 \\
%PBE$\alpha=2$-vdW & -341&-340 &-337 &-327  \\
B86-vdW     & $-328$&$ -327$&$ -324$& $-314$\\
%\hline
%devised
%PBEfix0.5-vdW & 24& 35& 26& 11& 117& 61\\
%old \PBEkappa-vdW &-332 &-332 &-329 &-320 \\
\PBEkappa-vdW &$-326$ &$-325$ &$-322$ &$-313$ \\
%rRPBE-vdW &20 &29 &22 &9 &99 &51 \\
%old optPBE-vdW &-340 &-340 &-338 &-329  \\
optPBE-vdW &$-335$ &$-334$ &$-332$ &$-323$  \\
optB88-vdW &$-347$ &$ -347$ &$ -344$ &$ -334 $  \\
optB88-vdW\footnote{Optimized self-consistently with GPAW} &$ -352$ & $-354$ &$ -349$ &$ -339$ \\
\hline\hline
\end{tabular}}
\end{table}

We have also applied the new functionals to another important class of problem, namely adsorption on surfaces. 
The accurate determination of adsorption energies is an issue of central importance to many disciplines.
However, in general, there is a
paucity of accurate reference data. 
Water on NaCl(001) is an exception where an adsorption energy 
of $-487$$\pm$60~meV at the $\Delta$CCSD(T) level was recently obtained using an embedded cluster approach~\cite{li2008}.
Using the geometry from ref.~\cite{li2008} and a slab model of the surface we computed adsorption energies for vdW-DF with revPBE exchange, B86 exchange
and the three new functionals. 
In contrast to revPBE-vdW, which yields an adsorption energy of $-334$~meV, the alternative
choices of exchange predict adsorption energies of $-413$ (\PBEkappa-vdW) to $-424$~meV (optB88).
Although the values presented are below the lower end of the error bar on the reference adsorption energy,
they are closer to it than revPBE-vdW and a regular GGA such as PBE ($-328$ meV).
% ~\cite{foot_lattice}.
%
This is an encouraging development with scope for improvement.

In summary, we have shown that the accuracy obtained from vdW-DF for a range of systems can be greatly improved
by making alternative choices for the GGA exchange component. Based on a combination of physical insight and optimization
three new exchange functionals have been proposed (\PBEkappa, optPBE, and optB88). 
\PBEkappa\ is the simplest alternative, optPBE is an optimized PBE-style functional that in 
addition to a low MAD on the S22 set yields
very precise results for the water hexamers, and optB88 yields the overall best performance on the S22 dataset.
%These functionals deliver
%very high precision from vdW-DF for the S22 dataset --- much better than chemical accuracy --- precise results for water clusters
%and improved performance over revPBE-vdW for water on salt. 
%These functionals represent only minor modifications of PBE and B88. They therefore come at no additional cost
%to regular revPBE based vdW calculations and are trivial to implement.
%
%If we compare the MADs obtained from other DFT-based vdW correction schemes (20~meV for ``LC-BOP+ALL''~\cite{sato2007}, 
%15~meV for B97-D~\cite{antony2006, grimme2007}, and 13~meV for the Tkatchenko and Scheffler method~\cite{tkatchenko2009}) 
%to those obtained here we find that the new functionals are comparable to or better than these other approaches.
%
%Let us now briefly compare our results to those obtained from other van der Waals correction schemes.
%If we consider the published MADs for the S22 dataset for the so-called LC-BOP+ALL (20~meV) method, 
%a method based on a range separated hybrid and a simpler version of vdW-DF~\cite{sato2007},
%B97-D (15~meV with a relatively small TZV(2df,2pd) basis set)~\cite{antony2006, grimme2007}, and a recently proposed
%algorithm from Tkatchenko and Scheffler (13~meV)~\cite{tkatchenko2009}, we see that our vdW functional-compatible
%exchange functionals yield results that are comparable or better than these approaches. 
%
%However, unlike the alternatives,
%vdW-DF does not require the application of a damping function which can affect performance and transferability.
%
We hope that this study lays the foundations for further improvements of vdW-DF and will enable more accurate 
treatments of a wide variety of dispersion and H bonded systems, such as liquid water.
%The accuracy exampled on the water clusters makes 
%this a good starting point to further improvements such as inclusion of exact exchange or changes of the vdW formula. 
Along with the 
recent efficiency improvements \cite{soler2009} we suggest that vdW-DF is now a serious medium-term contender for high precision simulations before more rigorous approaches such as the random phase approximation or quantum Monte Carlo become routine.

\begin{acknowledgments}

A. M. is supported by the EURYI scheme (see: www.esf.org/euryi), the EPSRC, and the European Research Council.
DRB is supported by the Royal Society.
We are very grateful to the London Centre for Nanotechnology, UCL Research Computing and the UK's HPC Materials Chemistry Consortium, which is funded by EPSRC (EP/F067496),
for computer time.

\end{acknowledgments}


\begin{thebibliography}{10}
\bibitem{new_xc_note}
The new exchange functionals reported that were not already in VASP were
  implemented in this study.
\bibitem{foot_dens}
{\rm For the entire S22 set the vdW corrections to the interaction energy are
  within 2~meV of each other, irrespective of whether B86, PBE, or revPBE
  density is used.}
\bibitem{geom_note}
The MAD in the O--O distances is only 0.01~\AA\ and in the H bond angles it is
  0.69$^\circ$ when the optB88-vdW optimized and reference geometries are
  compared.
\end{thebibliography}
\end{document}